\documentclass[12pt]{iopart}

\usepackage{fancyhdr}

\usepackage{iopams}
\usepackage{graphics,wrapfig,times}
\usepackage{graphicx}

\usepackage{epsfig,epstopdf}

\usepackage{amsfonts}
\usepackage[T1]{fontenc}

\def\mR{\mathcal{R}}

\def\be{\begin{equation}}
\def\ee{\end{equation}}
\def\bea{\begin{eqnarray}}
\def\eea{\end{eqnarray}}
\def\ba{\begin{array}}
\def\ea{\end{array}}
\def\w{\wedge}
\def\up{\stackrel}
\def\d{\textrm{d}}

\def\nn{\nonumber}

\begin{document}


\title[Inflation in $R + R^2$ Gravity with Torsion]{Inflation in $R + R^2$ Gravity with Torsion}

\author{Chih-Hung Wang$^1$ and Yu-Huei Wu$^{1,}$ $^2$}

\address{$^1$Department of Physics, National Central University,
     Chungli 320, Taiwan, R. O. C.}

\address{$^2$Center for Mathematics and Theoretical Physics, National Central University,  Chungli 320, Taiwan, R. O. C.}


\ead{$^1$\mailto{chwang@phy.ncu.edu.tw},
$^2$\mailto{yhwu@phy.ncu.edu.tw}}

\begin{abstract}

We examine an inflationary model in $R + R^2$ gravity with torsion, where $R^2$ denotes five independent quadratic curvature invariants; it turns out that only two free parameters remain in this model. We show that the behavior of the scale factor $a(t)$ is determined by two scalar fields, axial torsion $\chi(t)$ and the totally anti-symmetric curvature $E(t)$, which satisfy two first-order differential equations. Considering $\dot{\chi}\approx 0$ during inflation leads to a power-law inflation: $a \sim (t+ A)^p$ where $1< p \leq 2 $, and the constant $A$ is determined by the initial values of $E$, $\chi$ and the two parameters. After the end of inflation, $\chi$ and $E$ will enter into an oscillatory phase.

\end{abstract}

 \submitto{\CQG}

\maketitle

\section{Introduction}

Since cosmic microwave background radiation (CMB) was discovered, the standard hot big bang cosmological model becomes widely accepted. However, the hot big bang Universe leads to several problems, e.g. horizon and flatness problems \cite{liddellyth99}. Inflationary scenario was then proposed to resolve these problems (see \cite{linde05}). The standard approach to inflationary Universe is to introduce scalar inflaton fields without modifying general relativity (GR).  However, these inflaton fields will violate the strong energy condition. Instead of introducing inflatons, it has been found that several alternative theories of gravity can also provide inflationary Universe. A well-known pure gravity inflationary model is based on a fourth-order gravity theory $\hat{\mR} + \epsilon \hat{\mR}^2$, i.e., adding a quadratic pseudo-Riemannian scalar curvature $\hat{\mR}^2$ to the Einstein-Hibert action without introducing any scalar inflaton field \cite{starobinsky80,whitt84, mijic86}. In the following, we put hat on any geometric object $W$ to indicate that $\hat{W}$ is defined from Levi-Civita connection $\hat{\nabla}$.   In \cite{whitt84} (see \cite{wands94} for further generalization), it provides a better understanding of $\hat{\mR}^2$ inflation by showing that the Lagrangian $-2 \Lambda + \hat{\mR} + \epsilon \hat{\mR}^2$ is conformally related to Einstein gravity plus a scalar field, where $\Lambda$ denotes the cosmological constant. Furthermore, it has been demonstrated that the Lagrangian $-2 \Lambda + \hat{\mR} + \epsilon \hat{\mR}^2$ can be considered as a general quadratic curvature Lagrangian for homogeneous, isotropic cosmological model, i.e., pseudo-Riemannian curvature square $\hat{\mR}^{abcd} \hat{\mR}_{abcd}$ and Ricci curvature square $\hat{\mR}^{ab}\hat{\mR}_{ab}$ can be expressed in terms of $\hat{\mR}^2$ by using the Gauss-Bonnet theorem and a homogeneous, isotropic assumption \cite{barrow83}. It should be stressed that this demonstration will not be true in Riemann-Cartan spacetime, i.e., metric-compatible connection $\nabla g=0$ with torsion. 

Since magnitudes of curvature are expected to be very large in the early Universe, the effects of quadratic curvature invariants may dominate. So it is reasonable to include them in a Lagrangian,  which is a linear function of scalar curvautre. However,  \emph{it is not satisfactory to restrict these investigations in pseudo-Riemannian geometry, since the connection $\nabla$ and the metric $g$ are fundamentally independent}. Therefore, the fundamental variables of gravity should be considered as $\{g, \nabla\}$. It motivates us to investigate quadratic curvature effects, which is based on Riemann-Cartan spacetime, during inflationary epoch. In this paper, we extends the Einstein-Cartan theory to include all of quadratic curvature invariants, i.e. five independent invariants, in the early Universe.  Since dimension of torsion and curvature are $L^{-1}$ and $L^{-2}$ respectively, quadratic torsion effects can be neglected by comparing them to quadratic curvature in the early Universe.  So we do not include quadratic torsion terms in the Lagrangian. It turns out to be a degenerate case of the Poincar\'e gauge theory of Gravity (PGT), which does have three quadratic torsion invariants \cite{hehletal76}.   A homogeneous, isotropic cosmological model in the PGT has been developed and many special cases and solutions are discovered \cite{goennermuller81, minkevich06, minkevich07, shie_yo_nester08}. These special cases have been used to explain dark energy problems \cite{minkevich07, shie_yo_nester08} and inflation \cite{minkevich06}. In \cite{minkevich06}, the authors reduced the PGT field equations into a second-order differential equations with respect to Hubble radius by certain restrictions on indefinite parameters and also introduced a scalar field and matter fields during inflation. However, our inflationary model is completely different from ref \cite{minkevich06}. We start from a general quadratic curvature Lagrangian in the Riemann-Cartan spacetime and then obtain a pure gravity inflationary model without introducing any scalar or matter fields.

Our purpose in this paper is to investigate quadratic curvature effects in the early Universe and also to analyze whether these effects will generate inflation. Since our analysis is based on the Riemann-Cartan spacetime, there are five independent quadratic curvature invariants in the Lagrangian. It is worth to point out that our work is not a special case of $f({\mR})$ gravity with torsion \cite{capozzielloetal07}, where $\mR$ is a scalar curvature. In the $f(\mR)$ gravity with torsion, it does not include all possible curvature invariants, e.g. Ricci square $\mR^{ab} \mR_{ab}$, in its Lagrangian. There is no fundamental reason for us to neglect these terms in the early Universe. Moreover, we do find that these neglected terms can yield a power-law inflation. The field equations of $f(\mR)$ gravity with torsion give totally antisymmetric torsion tensor, i.e. axial-vector torsion, vanishing. Conversely, our inflationary solution is purely determined by the time-component of axial-vector torsion $\chi$ called axial torsion.  

The plan of this paper is as follows. In Section 2, we give a brief review of Riemann-Cartan geometry. Moreover, we start from a general quadratic curvature Lagrangian and obtain an effective quadratic curvature Lagrangian, which contain only three quadratic curvature invariants, for a homogeneous, isotropic cosmological model. A similar argument has been done in pseudo-Riemannian spacetime \cite{barrow83, mijic86}. In Section 3, we derive the equations of motion for inflationary cosmology. Since we assume that there is no matter field during inflation, it further reduces one more parameter. The equations of motion will contain only two parameters. In Section 4, we find an inflationary solution which are corresponding to a power-law inflation. The e-folding number $N$, which is determined by one of two parameters, can satisfy the standard requirement $N > 60$. At the end of inflation, $\chi$ will enter into an oscillatory phase. In this paper, $\hbar=c=1$ and the metric signature is $(- + + +)$.


\section{Riemann-Cartan geometry and the quadratic curvature Lagrangian}

Since Einstein developed GR in terms of a pseudo-Riemannian spacetime, it was argued that GR has a pre-geometry, especially when one discovered that intrinsic spins can naturally generate torsion tensor.  A directly extended theory of GR to include torsion is called Einsten-Cartan theory, which is based on a Riemann-Cartan spacetime.  If the spin densities of matter fields vanish, the Einstein-Cartan theory will return to GR. Since torsion effects generated by spin sources is too small, the current experiments still cannot distinguish their differences.  However, these two theories may become significantly different in early Universe due to  quadratic curvature effects.

The geometry of a Riemann-Cartan spacetime is described by metric $g$ and metric-compatible connection with torsion $\nabla$. The curvature 2-forms $R^a{_b}$ and torsion 2-forms $T^a$ can be expressed in terms of connection 1-forms $\omega^a{_b}$ and an orthonormal co-frame $e^a$:
\bea
R^a{_b} &=& \d \omega^a{_b} + \omega^a{_c} \w \omega^c{_b},\\
T^a &=& \d e^a + \omega^a{_b} \w e^b=D e^a, 
\eea where $D$ is the covariant exterior derivative \cite{benntucker87}. In an orthonormal frame, $\omega_{ab}$ satisfy antisymmetric condition 
\bea
\omega_{ab} + \omega_{ba}=0,
\eea where the indices are lowered by $\eta_{ab}=\textrm{diag} (-1, 1, 1, 1)$. In the following, Latin indices are raised and lowered by $\eta_{ab}$.

It has been shown that the curvature 2-forms $R_{ab}$ can be irreducibly decomposed into the following pieces \cite{mccrea92} : $R_{ab} = \up{1}{R}_{ab} \textrm{(Weyl)} + \up{2}{R}_{ab} (\textrm{Paircom}) +\up{3}{R}_{ab} (\textrm{Pseudoscalar}) +\up{4}{R}_{ab} (\textrm{SymRicci}) +  \up{5}{R}_{ab} (\textrm{AntiRicci}) + \up{6}{R}_{ab} (\textrm{Scalar})$. Moreover, the irreducible components of torsion 2-forms yield $T^a = \up{1}{T^a} (\textrm{Tensor}) + \up{2}{T^a} (\textrm{Vector}) + \up{3}{T^a} (\textrm{Axitor}) $. It should be pointed out that one has $\up{2}{R}_{ab}=\up{3}{R}_{ab}=\up{5}{R}_{ab} \equiv 0$ and $T^a\equiv 0$ in the pseudo-Riemannian spacetime. In terms of $\up{\alpha}{R}_{ab}$, the most general quadratic curvature action (neglected any possible derivative on curvature) in the Riemann-Cartan spacetime is given by 
\be
S[e^a, \,\omega^a{_b}]=\frac{1}{2\kappa}\int_M [\, \Lambda *1 + c_0\,R_{ab} \w * e^{ab}  + R^{ab} \w \sum^6_{\alpha=1} c_\alpha  \,*\up{\alpha}{R}_{ab}\,], \label{action}
\ee where $\kappa=8\pi G$ and $e^{a\cdots b}{_{c\cdots d}}$ denotes $e^a \w \cdots \w e^b \w e_c \w \cdots \w e_d$. $\Lambda$ is the cosmological constant and $*$ denotes the Hodge map associated with $g$. $c_0$ is a dimensionless parameter and $[c_\alpha]=L^{2}$. In the PGT, the most general gauge invariant action should also includes three irreducible pieces of torsion square in its Lagrangian \cite{hehletal76, goennermuller81}. Although $R_{ab}$ have six irreducible components, there are only five independent quadratic curvature invariants due to generalized Gauss-Bonnet theorem, which gives \cite{nieh80} 
\bea
 \varepsilon_{abcd} R^{ab} \w R^{cd} = \textrm{an exact form}. \label{gb}
\eea Using (\ref{gb}), one of irreducible components of $R_{ab}$ can be replaced in terms of an exact form which does not contribute to field equations.  So it turns out to have only six parameters (including $c_0$) in the field equations, which are obtained from varying (\ref{action}).  In order to obtain the Newtonian theory in weak field approximation \cite{hayashi80, muller-hoissen83}, we should fix $c_0=1$, which actually gives Einstein-Cartan action with $\Lambda$ if quadratic curvature invariants vanish.

It is straightforward to obtain field equations by varying $S$ with respect to $\{e^a\}$ and $\{\omega_{ab}\}$ and these equations should contain five independent parameters.\footnote{These field equations are equivalent to the field equations (3.3) and (3.4) in \cite{goennermuller81} by neglecting torsion square terms and rearranging parameters of quadratic curvature invariants in (3.3)-(3.4).} However, one can further reduce two parameters in the field equations according to an assumption of homogeneous, isotropic Universe.  It can be verified that the homogeneous and isotropic assumption yields $\up{1}{R}_{ab} = \up{5}{R}_{ab}=0$ by using (\ref{g}) and (\ref{T^a}), so quadratic $\up{1}{R}_{ab}$ and $\up{5}{R}_{ab}$ do not contribute to field equations for homogeneous, isotropic cosmology. It is worth to point out that they may participate in the \emph{perturbed} field equations.\footnote{The cosmological perturbation analysis for $R + R^2$ gravity will relate to our future works.}  Since our current work is only concentrated on homogeneous and isotropic Universe, we can start from the following effective action 
\bea
S_{E}[e^a, \,\omega^a{_b}]=\frac{1}{2\kappa}&\int_M &  [  \Lambda *1 + \,R_{ab} \w * e^{ab} + b_3\, R_{ab} \w e^{ab} \w * ( R_{ab} \w e^{ab})] \nn\\
&&+ b_4\, P_a \w  * P^a  -  b_6\, R_{ab} \w * e^{ab} \w * ( R_{ab} \w * e^{ab}) \,], 
\eea  where $P_b= i_a R^a{_b}$ denotes Ricci 1-forms. So it only involves three indefinite parameters. In the following, we set $\Lambda=0$  since the effects of $\Lambda$ in the early Universe are expected to be small.

 Varying $S_E + S_{\rm{matter}}$ with respect to the orthonormal co-frames $\{e^a\}$ and connection 1-forms $\{\omega_{ab}\}$ gives following field equations:
\bea
&& R_{ab} \w * e^{ab}{_c} - b_3 \,E (E * e_c + 2\, i_c R_{ab} \w e^{ab} ) + b_4(\,i_c (P^b \w *P_b) \nn\\
&&+2\, i_c R^{ab} \w i_a *P_b)  + \,b_6\,\mathcal{R} (2 R_{ab} \w *e^{ab}{_c} - \mathcal{R} * e_c) = - 2\kappa\, \tau_c, \label{feq1}\\
 &&T^c \w *e^{ab}{_c} + 2\, b_3 D ( E\, e^{ab} ) - b_4 D*(P^b \w e^a - P^a \w e^b) \nn\\
 &&+ 2\,b_6 D (\mathcal{R} * e^{ab}) = - 2\kappa \,S^{ab}, \label{feq2}
 \eea where  $\mathcal{R}= i_a P^a$ is the scalar curvature and $E=\frac{1}{2} R_{abcd} \varepsilon^{abcd}$ is the totally antisymmetric curvature. $i_a$ denotes the interior derivative and $\ \varepsilon^{abcd}$ is the Levi-Civita antisymmetric $\varepsilon$-symbol. $\tau_c$ and $S^{ab}$ are stress-energy and spin density 3-forms of the matter fields.
 
\section{Equations of motion for inflationary cosmology}
 
In an isotropic, homogeneous cosmological model, the spacetime metric and torsion 2-forms become \cite{goennermuller81}:
\bea
g &=& - \d t \otimes \d t + a^2(t) \,(1+ \frac{1}{4}k r^2)^{-2}\,\sum^3_{A=1} \d x^A \otimes \d x^A,\label{g}\\
T^a &=& f(t) \,e^a \w e^0 + 2\, \chi(t) * (e^a \w e^0), \label{T^a}
\eea where curvature constant $k$= -1, 0, or 1. For simplicity, we only concentrate on the flat Universe $k=0$. Moreover, the non-vanishing components of $\tau_a$ and $S^{ab}$ yield
\bea
\tau_a&=& \eta_{0a} (\,\rho(t) + p(t)\,) * e_0 + p(t) * e_a, \label{tau_a}\\ 
S^{ab} &=& -q(t)\, e^{[a} \w * ( e^{b]} \w e^0) +  s(t) \,e^0 \w e^a \w e^b, \label{S^ab}
\eea where square bracket indicates index anti-symmetrisation.  In standard inflation models, the energy density $\rho$ and pressure $p$ of an inflaton field are given by $\rho=\frac{1}{2} \dot{\phi}^2 +V(\phi)$ and $p=\frac{1}{2} \dot{\phi}^2 - V(\phi)$ \cite{liddellyth99}. It still lacks a fundamental theory which can explain the origin of the inflaton fields and also derive the potential $V(\phi)$. Moreover, there is no observational evidence to yield any feature of matter fields during inflationary epoch.  
In this paper, we will assume $\tau_a=S^{ab}=0$ in the early Universe and then obtain a pure gravity inflationary model in the Riemann-Cartan spacetime.

Using (\ref{g})-(\ref{S^ab}), the field equations (\ref{feq1})-(\ref{feq2}) with $k=0$ yield\footnote{ (\ref{feq3})-(\ref{feq6}) is a degenerated case of the field equations (4.2)-(4.5) in \cite{goennermuller81}.}
\bea
&& \{(H + f)^2 - \chi^2\} + \frac{b_3}{6} E^2 - 4 b_3 E \chi ( H + f ) \nn\\
&& - \frac{(b_4 + 3 b_6)}{18} \mathcal{R}^2 + \frac{2 (b_4 + 3 b_6)}{3} \, \mathcal{R} \{(H + f)^2 - \chi^2\} =\frac{\kappa \rho}{3}, \label{feq3}\\
&&\mathcal{R}= \kappa (\rho - 3 p), \label{feq4}\\
&&(b_6 + \frac{b_4}{3}) (\dot{\mathcal{R}} - 2 \mathcal{R} f) + 2 b_4 \chi (H \chi + \dot{\chi}) + 2 b_3 E \chi -  f= - \frac{\kappa q}{2}, \label{feq5}\\
&& b_3 (\dot{E} - 2 f E) - \frac{( b_4 + 6\, b_6)}{3}  \mathcal{R} \chi - 2 b_4 \chi \{(H + f)^2 - \chi^2\} - \chi =- \kappa s, \label{feq6}
\eea where
\bea
&& \mathcal{R}= 6 (\dot{H} + 2 H^2 + 3 H f + \dot{f} + f^2 - \chi^2),\\
&&E= 6 (\dot{\chi} + 3 H \chi + 2 \chi f), \label{E}
\eea and $H=\frac{\dot{a}}{a}$ denotes the Hubble parameter. It can be shown that (\ref{feq3})-(\ref{feq6})  form a complete system to describe the evolution of the isotropic, homogeneous flat Universe once the equation of state $p=p(\rho)$ is given. In particular, $\rho= p=0$ (inflationary epoch) or $p=\frac{\rho}{3}$ (radiation domination) both yield $\mathcal{R}=0$ (see (\ref{feq4})). It is not difficult to see that (\ref{feq3}) yields the Friedmann equations when the quadratic curvature terms and the spin density vanish.  By substituting $\mathcal{R}=0$ into (\ref{feq3}), (\ref{feq5}) and (\ref{feq6}), one can easily verify that only $b_3$ and $b_4$ left in these equations.  

Since we consider $\rho=p=q=s=0$ during inflation, (\ref{feq3})-(\ref{feq6}) with (\ref{E}) then become  
\bea
\dot{E} &=& \frac{\chi}{b_3} + 2\, (\Phi - H ) E + \frac{2b_4}{b_3} \chi \,( \Phi^2 - \chi^2 ),  \label{feq7}\\
\dot{\chi}&=& \frac{E}{6} -  H \chi - 2 \chi\, \Phi, \label{feq8}
\eea with two algebraic equations:
\bea
\Phi&=& 2 b_3 E \chi \pm \mathcal{A}, \label{feq9}\\
H &=& - \frac{b_4}{3} E \chi + 8 b_3\, b_4\, E \chi^3 \pm (1 + 4 b_4 \chi^2) \mathcal{A}, \label{feq10}
\eea where $\mathcal{A} = \sqrt{ (2 b_3 E \chi )^2 + \chi^2 - \frac{b_3 E^2}{6}}$ and $\Phi= H + f$. We further simplify (\ref{feq7}) and (\ref{feq8}) by substituting (\ref{feq9}) and (\ref{feq10}) into them, which gives
\bea
\dot{E} &=& \frac{\chi}{b_3}  + (4 b_3 + \frac{b_4}{3})\,E^2 \chi, \label{feq11}\\
\dot{\chi}&=& \frac{E}{6} + (\frac{b_4}{3} - 4 b_3) E \chi^2 \mp (3 + 4 b_4 \chi^2 ) \mathcal{A} \chi - 8 b_3 b_4 E \chi^4, \label{feq12}
\eea It turns out that the evolution of the inflationary Universe is characterized by $b_3$, $b_4$  and the initial values ($t=0$) of $E$ and $\chi$. We should point out that here $t=0$ denotes the beginning of inflation. It is reasonable to require $b_3 \equiv - b < 0$ \footnote{The requirement of positive kinetic energy in the spin $0^-$  mode of linearized PGT also gives $b_3 <0$ \cite{hayashi80}. } in order to ensure that $\mathcal{A}$ is a real function. Moreover, we assume that the initial values $E_0$ and $\chi_0$ are both less than or on the order of the Planck scale to ensure the classical validity of the evolution \cite{mijic86}. 

\section{Power-law inflation}
We first notice that (\ref{feq10})-(\ref{feq12}) has a similarity to hybrid inflation models \cite{linde90,linde91}, with the effective potential given by $V(E, \chi)= 3 \kappa^{-1} H^2$.  In hybrid inflation models, the effective potentials $V(\sigma, \phi)$ are constructed in such a way that one scalar field $\sigma=0$ during slow-roll inflation, and starts to roll down to its true vacuum when the other scalar field $\phi$ falls to a critical value $\phi_c$. We adopt a similar arguement by considering $| \chi | \approx | \chi_0 | \ll  l_{\rm{ph}}^{-1}$ during inflation and $| E_0 | \sim l_{\rm{ph}}^{-2}$. By neglecting $\chi^2$ in $\mathcal{A}$,  (\ref{feq12}) reduces to an algebraic cubic equation
\be
 (32\alpha + \frac{8 \alpha^2}{3} ) \,\bar{\chi}^3 - (20  + \frac{4\alpha}{3}  - \frac{\alpha^2}{9} )\,\bar{\chi}^2 - (\frac{1}{6} + \frac{\alpha}{9})\, \bar{\chi} + \frac{1}{36} \approx 0, \label{feq13}
\ee where $\bar{\chi} = b \chi_0^2 >0$ and $\alpha= - \frac{b_4}{b}$, which are both dimensionless parameters. From numerically and analytically analyzing (\ref{feq13}), we find that the condition $\alpha \sim \bar{\chi} \sim O(1)$, which is consistent with neglecting $\chi^2$ in $\mathcal{A}$, can naturally provide the sufficient number of e-folding (see below), so we will assume the condition being satisfying in the following discussions. Since $b=\frac{\bar{\chi}}{\chi_0^2}\gg l_{\rm{ph}}^2$,
one can neglect $\frac{\chi}{b_3}$ in (\ref{feq11}), and it becomes
\be
\dot{E} \approx - (4 + \frac{\alpha}{3}) \,b \chi_0 E^2. \label{feq14}
\ee  (\ref{feq14}) then yields a general solution
\be
E\approx \mathcal{B} \left(t + \frac{\mathcal{B}}{E_0} \right)^{-1}, \label{solutionE}
\ee where $\mathcal{B}= \frac{3}{(12 + \alpha)\, b \chi_0}$. 
By substituting (\ref{solutionE}) into (\ref{feq10}), we obtain
 $H\approx p\, (t + \frac{\mathcal{B}}{E_0})^{-1} $, where 
\be
p=\frac{3}{12 + \alpha} \left(\frac{\alpha}{3} + 8 \alpha \bar{\chi} \pm (1- 4 \alpha \bar{\chi})\sqrt{4 + \frac{1}{6 \bar{\chi}}}\right) \label{p}
\ee
for the case $\chi_0 E >0$, and $\pm$ should be changed to $\mp$ for the other case $\chi_0 E < 0$. So a solution of $a(t)$ for $a(t)>0$ gives
\be
a \approx A_0\,(t + \frac{\mathcal{B}}{E_0})^p, \label{solutiona}
\ee where $A_0=  a_0 (\frac{\mathcal{B}}{E_0})^{-p} >0$, and $a_0>0$ denotes the initial value of $a$. 

It is known that a general criterion for inflation is $\ddot{a} > 0$, which may be sufficient to solve the horizon and flatness problems. It turns out that to require $p>1$ in  (\ref{solutiona}) corresponds a new type of the power-law inflation \cite{abbottwise84,lucchin85}.  From (\ref{p}), one can see that $p>1$ will limit the value of $\alpha$. For example, if we consider a degenerate case $\alpha=0$, it gives $p=\frac{3}{4}$ which does not have inflation. However, in the $\alpha=1$ case, which yields $\bar{\chi} \approx 0.623$, the numerical calculations of (\ref{p}) while choosing the negative sign in the round bracket yields $p\approx 1.938$, which gives power-law inflation. From numerical calculations, we also find that $1\leq p \leq 2$ when $\alpha> 0$. In addition to $\ddot{a} >0$, the amount of inflation is usually required to satisfy $N>60$, where $N\equiv\ln\frac{a(t_{\rm{end}})}{a_0}$ denotes the number of e-foldings, and $t_{\rm{end}}$ means the end of inflation. In order to estimate $N$, we consider that the inflation comes to an end when (\ref{feq13}) is no longer valid, and it occurs when $\chi^2$ in $\mathcal{A}$ cannot be neglected , i.e. 
\be
| b\,E^2_{\rm{end}} | \sim \chi_0^2 \hspace{1cm}\textrm{or}\hspace{1cm}  | b^2 E^2_{\rm{end}} | \sim O(1). \label{endE}
\ee  From (\ref{solutionE}) and (\ref{endE}) with the help of $\alpha \sim \bar{\chi} \sim O(1)$, the values $E_{\rm{end}}$ and $t_{\rm{end}}$ can be estimated as $E_{\rm{end}} \approx \frac{3}{(12 + \alpha)\, b}$ and $t_{\rm{end}} \approx \frac{1}{\chi_0}$. By substituting $t_{\rm{end}} \approx \frac{1}{\chi_0}$ into the number of e-foldings
\be
N = \ln \left(\frac{E_0}{\mathcal{B}}\, t_{\rm{end}}+ 1\right)^p,
\ee we obtain $e^N \approx ( \,(4 +\frac{\alpha}{3})E_0\,b)^p$. It turns out that the value of $N$ is determined by the parameter $b$. In the $\alpha=1$ case, $N>60$ requires $b > 10^{15}\, l_{\rm{ph}}^{2} \sim 10^{-55}\, m^{2}$, which satisfies the tests of solar experiments \cite{gladchenko94}. After the end of inflation, $\chi$ and $E$  both become very small, and the linear terms in (\ref{feq7}) and (\ref{feq8}) should finally dominate. In the linearized equations of (\ref{feq11}) and (\ref{feq12}), one can easily obtain oscillatory solutions: $\chi \sim \sin \frac{t}{\sqrt{6\, b}}$ and $E \sim  \cos \frac{t}{\sqrt{6\, b}}$. Whether these oscillatory phases can provide a reheating process is still in progress. 

\section{Conclusion}

In conclusion, we first discovered a pure gravity inflationary model in $R + R^2$ gravity theories with torsion, and presented a detail analytic analysis of this model. It turns out that the dynamics of the inflationary model is purely determined by two parameters $b$ and $\alpha$ with two initial values $E_0$ and $\chi_0$.  We further found that $ | \chi_0 | \ll  l_{\rm{ph}}^{-1}$ and $| E_0 | \sim l_{\rm{ph}}^{-2}$ yields a power-law inflation if $\alpha >0$. In this power-law inflation solution, $\chi$ is near a constant $\chi_0$ and $E\sim t^{-1}$. Since $\chi$ cannot be near $\chi_0$ after $E$ decays to satisfy the condition $| b\,E^2_{\rm{end}} | \sim \chi_0^2$, it can be considered as the end of inflation. The period of inflation can be estimated to be  $\frac{1}{\chi_0}$.     Moreover, the e-folding number $N$ is completely determined by $b$ and also satisfy the requirement $N>60$ when $b > 10^{15}\, l_{\rm{ph}}^{2} \sim 10^{-55}\, m^{2}$. 

After the end of inflation, the system may enter into an asymptotical regiem which yields an oscillatory phases. These oscillatory phases are expected to give a reheating process. Since CMB anisotropy has been discovered, a further understanding of the reheating process and primordial perturbations will provide more restricted constraints on $b$, $\alpha$, $E_0$ and $\chi_0$. These issues will be considered as our future work.

\ack
CHW would like to thank Prof James M. Nester and Prof Hwei-Jang Yo for helpful discussions. This work was supported by National Science Council of R.O.C. under Grant No. NSC 096-2811-M-008-040 (CHW).


\section*{References}

\begin{thebibliography}{20}

\bibitem{abbottwise84}
Abbott L F and Wise M B, 1984 \NP B {\bf 244}  541.

\bibitem{barrow83}
 Barrow J D and Ottewill A C, 1983 \JPA  {\bf 16}  2757.

\bibitem{benntucker87}
 Benn I M and Tucker R W, 1987 {\it An introduction to spinors and geometry with applications to physics} (Bristol: Institute of Physics Publishing)

 
\bibitem{capozzielloetal07}
Capozziello S, Cianci R, Stornaiolo C and Vignolo S, 2007 \CQG {\bf 24}  6417.

\bibitem{hayashi80}
 Hayashi K and Shirafuji T, 1980 {\it Prog. Theor. Phys.} {\bf 64}  866, 883, 1435, 2222.

\bibitem{hehletal76}
Hehl F W,  von der Heyde P,  Kerlick G D and Nester J, 1976 \RMP {\bf 48}  393.

\bibitem{gladchenko94}
Gladchenko M S and  Zhytnikov V V, 1994 \PR D {\bf 50}  5060.

\bibitem{goennermuller81}
Goenner H and  M\"uller-Hoissen F, 1984 \CQG {\bf 1}  651.

\bibitem{liddellyth99}
 Liddel A R and  Lyth D, 2000 {\it Cosmological Inflation and Large-Scale Structure} (Cambridge: Cambridge University Press).

\bibitem{linde05}
Linde A D, 1990 {\it Particle Physics and Inflationary Cosmology} (Harwood, Chur, Switzerland)

\bibitem{linde90}
 Linde A D,  1990 \PL B {\bf 249}  18.

\bibitem{linde91}
 Linde A D,  1991 \PL\ B {\bf 259}  38.

\bibitem{lucchin85}
 Lucchin F and Matarrese S, 1985 \PR D {\bf 32}  1316.
 
 \bibitem{mccrea92}
 McCrea J D, 1992 \CQG {\bf 9}  553.

\bibitem{mijic86}
 Miji\'c M B,  Morris M M and  Suen W -M, 1986 \PR D {\bf 34}  2934.

\bibitem{minkevich06}
 Minkevich A V and Garkun A S 2006 \CQG {\bf 23}  4237. 
 
\bibitem{minkevich07}
 Minkevich A V, Garkun A S and Kudin V I 2007 \CQG {\bf 24}  5835. 

\bibitem{muller-hoissen83}
 M\"uller-Hoissen F, 1983 {\it Gen. Rel. Grav.} {\bf 15}  1051.

\bibitem{nieh80}
 Nieh H T, 1980 \JMP {\bf 21} 1439.
 
\bibitem{shie_yo_nester08}
 Shie K F,  Nester J M and  Yo H J, 2008 \PR D {\bf 78}  023522.

\bibitem{starobinsky80}
Starobinsky A A, 1980 \PL  B {\bf 91}  99.

\bibitem{wands94}
Wands D, 1994 \CQG {\bf 11}  269.

\bibitem{whitt84}
 Whitt B, 1984 \PL B {\bf 145}  176.


\end{thebibliography}
\end{document}